\documentclass[pre,twocolumn,preprintnumbers,amsmath,amssymb,showpacs,
nofootinbib,floatfix]{revtex4}
\usepackage{graphicx,bm,amsmath}

\def\<{\langle}
\def\>{\rangle}

\begin{document}
\title{Universality of the glassy transitions in 
the two-dimensional $\pm J$ Ising model}
\author{Francesco Parisen Toldin}
\email{parisen@mf.mpg.de}
\affiliation{Max-Planck-Institut f\"ur Metallforschung,
Heisenbergstrasse 3, D-70569 Stuttgart, Germany
}\affiliation{
Institut f\"ur Theoretische und Angewandte Physik,
Universit\"at Stuttgart,
Pfaffenwaldring 57,
D-70569 Stuttgart,
Germany
}
\author{Andrea Pelissetto}
\email{Andrea.Pelissetto@roma1.infn.it}
\affiliation{
Dipartimento di Fisica dell'Universit\`a di Roma ``La Sapienza'' and INFN,
Piazzale Aldo Moro 2,
I-00185 Roma,
Italy
}
\author{Ettore Vicari}
\email{vicari@df.unipi.it}
\affiliation{
Dipartimento di Fisica dell'Universit\`a di Pisa and INFN,
Largo Pontecorvo 3,
I-56127 Pisa,
Italy
}

\begin{abstract}
  We investigate the zero-temperature glassy transitions
  in the square-lattice $\pm J$ Ising
  model, with bond distribution $P(J_{xy}) = p \delta(J_{xy} - J) + (1-p)
  \delta(J_{xy} + J)$; $p=1$ and $p=1/2$ correspond to the pure
  Ising model and to the Ising spin glass with symmetric bimodal distribution,
  respectively.
  We present finite-temperature Monte Carlo simulations at $p=4/5$, which is
  close to the low-temperature paramagnetic-ferromagnetic transition line
  located at $p\approx 0.89$, and at $p=1/2$.  Their comparison provides a
  strong evidence that the glassy critical behavior that occurs for 
  $1-p_0<p<p_0$, $p_0\approx 0.897$, is universal, i.e., independent of $p$.
  Moreover, we show that glassy and magnetic modes are 
  not coupled at the multicritical zero-temperature point
  where the paramagnetic-ferromagnetic transition line and the 
  $T=0$ glassy transition line meet. On the theoretical side we discuss the 
  validity of finite-size scaling in glassy systems with a zero-temperature 
  transition and a discrete Hamiltonian spectrum. Because of a freezing 
  phenomenon which occurs in a finite volume at sufficiently low temperatures,
  the standard finite-size scaling limit in terms of $TL^{1/\nu}$ does not 
  exist: the renormalization-group invariant quantity $\xi/L$ should be used 
  instead as basic variable.
\end{abstract}


\maketitle

\section{Introduction}
\label{sec:intro}

The $\pm J$ Ising model~\cite{EA-75} is a standard theoretical laboratory to
study the effects of quenched disorder and frustration on
the critical behavior of spin systems.  We consider the two-dimensional (2D)
$\pm J$ Ising model defined on a square lattice by the Hamiltonian
\begin{equation}
{\cal H} = - \sum_{\langle xy \rangle} J_{xy} \sigma_x \sigma_y,
\label{lH}
\end{equation}
where $\sigma_x=\pm 1$, the sum is over all pairs of lattice nearest-neighbor
sites, and the exchange interactions $J_{xy}$ are uncorrelated quenched random
variables, taking values $\pm J$ with probability distribution
\begin{equation}
P(J_{xy}) = p \delta(J_{xy} - J) + (1-p) \delta(J_{xy} + J). 
\label{probdis}
\end{equation}
For $p=1$ we recover the standard Ising model, while for $p=1/2$ we obtain the
so-called bimodal Ising glass model.  For $p\neq 1/2$, the disorder average of
the couplings is given by $[J_{xy}]=J(2p-1)\neq 0$ and ferromagnetic (or
antiferromagnetic) configurations are energetically favored.  Note that its
thermodynamic behavior is symmetric under $p\rightarrow 1-p$.  In the
following we set $J=1$ without loss of generality.

\begin{figure}
\includegraphics[width=20em,keepaspectratio]{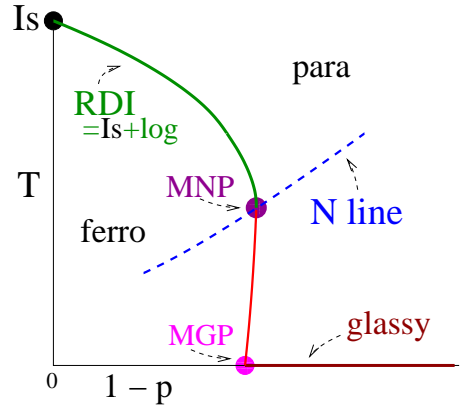}
\caption{Phase diagram of the square-lattice $\pm J$ Ising model
in the $T$-$p$ plane. 
The phase diagram is symmetric under $p\rightarrow 1-p$.}
\label{phasedia}
\end{figure}

The 2D $\pm J$ Ising model has been extensively investigated; see, e.g.,
Refs.~\cite{Nishimori-01,KR-04,PTPV-09} for recent reviews. 
As sketched in Fig.~\ref{phasedia},
at finite temperature it presents 
a paramagnetic and a ferromagnetic phase.  
They are separated by a transition line, which starts at
the pure Ising transition point, at $p=1$ and $T_{\rm
  Is}=2/\ln(1+\sqrt{2})=2.26919...$, and ends at the disorder-driven
ferromagnetic $T=0$ transition, at~\cite{WHP-03,AH-04} $p_0\approx 0.897$.
The point where this transition line meets the so-called Nishimori (N)
line~\cite{Nishimori-81,Nishimori-01,Nline}, at~\cite{PTPV-09} $T_M=0.9527(1)$
and $p_M=0.89083(3)$ (see also Refs.~\cite{NN-02,Ohzeki-08} for analytical
estimates of $T_M,p_M$) is a multicritical point (MNP)~\cite{LH-89,HPTPV-08}.
The MNP divides the paramagnetic-ferromagnetic (PF) transition line in two
parts.  The PF transition line from the Ising point at $p=1$ to the MNP is
controlled by the Ising fixed point. Here disorder gives only rise to
(universal) logarithmic corrections to the standard Ising critical behavior;
see, e.g., Ref.~\cite{HPTPV-08b} and references therein.  The slightly
reentrant low-temperature PF transition for $T<T_M$ belongs instead 
to a different
strong-disorder (SDI) universality class~\cite{McMillan-84,MB-98,PTPV-09}.

Several studies
\cite{RSBDJR-96,Houdayer-01,HY-01,CBM-02,AMMP-03,LGMMR-04,HH-04,
  KLY-04,KL-05,JLMM-06,KLC-07,Hartmann-08,ON-09} have discussed the possible
existence of a glassy transition, considering in most of the cases models with
symmetric disorder distributions such that $[J_{xy}]=0$.  At variance with the
three-dimensional case, no finite-temperature glassy phase occurs and a
critical behavior is only observed at $T=0$.  Moreover, recent results for the
bimodal Ising model \cite{JLMM-06,KLC-07,Hartmann-08,ON-09} have provided
compelling evidence that the bimodal Ising model and other models with
symmetric continuous disorder distributions, for instance the Gaussian
distribution, undergo a zero-temperature glassy transition in the same
universality class.  In all cases, for $T\to 0$ the correlation length
increases as $T^{-\nu}$ with \cite{footnote-nu} $\nu\approx 3.55$.  Even
though the asymptotic critical behavior is the same, for small values of the
temperature and in finite volume the bimodal Ising glass model, and in general
any model with a discrete disorder distribution, presents a peculiar
phenomenon related to the discreteness of the Hamiltonian spectrum.  This
implies that, although the bimodal and Gaussian Ising glass have the same
critical behavior in thermodynamic limit~\cite{JLMM-06}, i.e., if one takes
$L\to\infty$ before $T\to 0$, much more care has to be exercized in the
finite-size scaling (FSS) limit. In particular, as we shall see, in the
bimodal Ising glass model one cannot observe the standard FSS limit in terms
of the scaling variable $TL^{1/\nu}$. An appropriate variable is instead the
renormalization-group (RG) invariant quantity $\xi/L$.

In the case of the 2D $\pm J$ Ising model, a natural scenario is that 
a zero-temperature glassy transition occurs for any $p$ in the range 
$1-p_0<p<p_0$, and that the
glassy critical behavior is independent of $p$. This implies that a nonzero
$[J_{xy}]$ is irrelevant for the critical behavior, as found in mean-field
models~\cite{Toulouse-80} and in the 3D $\pm J$ Ising model~\cite{HPV-08}.

In this paper, which completes a series of
papers~\cite{PTPV-09,HPTPV-08,HPTPV-08b} devoted to the study of the phase
diagram and critical behavior of the 2D $\pm J$ Ising model, we investigate
the glassy behavior for $1-p_0<p<p_0$.  For this purpose, we present Monte
Carlo (MC) simulations at $p=4/5$, which is relatively close to the
low-temperature PF line ($p_0\approx p_{MNP} \approx 0.89$), and at $p=1/2$,
up to lattice sizes $L=64$ and for $T\gtrsim 0.1$. As we shall see, our
results provide a strong evidence of the universality of the glassy
zero-temperature critical behavior and thus provide strong support to the
scenario of a universal glassy critical line for $T=0$ and $1-p_0<p<p_0$.
Moreover, we provide evidence that the magnetic and glassy behaviors at the
$T=0$ multicritical glassy point (MGP), where the low-temperature PF and the
$T=0$ glassy transition lines meet, at $p_0\approx 0.897$, see
Fig.~\ref{phasedia}, are decoupled. Finally, we discuss the critical behavior
of the overlap quantities along the PF line that connects the MNP to the $T=0$
MGP: we observe an apparently $T$-dependent critical behavior.

The paper is organized as follows. In Sec.~\ref{sec:mc} we define the
quantities we have considered in the MC simulations.  In
Sec.~\ref{sec:results} we discuss the behavior at the $T=0$ glassy transition.
In particular, we discuss the freezing phenomenon observed in a finite volume
at very small temperatures due to the discreteness of the Hamiltonian
spectrum, the universality of the glassy critical behavior, and the critical
behavior of the overlap susceptibility.  In Sec.~\ref{sec:resultsPF} we
discuss the critical behavior of the overlap correlations along the
low-temperature paramagnetic-ferromagnetic transition line, below the MNP.
Finally, In Sec.~\ref{sec:conclusions} we present our conclusions.

\section{Definitions}
\label{sec:mc}

The critical modes at the glassy transition are those related to the overlap
variable $q_x \equiv \sigma_x^{(1)} \sigma_x^{(2)}$, where the spins
$\sigma_x^{(i)}$ belong to two independent replicas with the same disorder
realization $\{J_{xy}\}$.  In our MC simulations we measure the overlap
susceptibility $\chi$ and the second-moment correlation length $\xi$ defined
from the correlation function $G_o(x) \equiv [\langle q_0 \,q_x \rangle] = 
[\langle \sigma_0 \,\sigma_x \rangle^2]$, 
where the
angular and the square brackets indicate the thermal average and the quenched
average over disorder, respectively. We define 
$\chi \equiv \sum_{x} G_o(x)$ and
\begin{eqnarray}
\xi^2 \equiv  {1\over 4 \sin^2 (p_{\rm min}/2)} 
{\widetilde{G}_o(0) - \widetilde{G}_o(p)\over \widetilde{G}_o(p)},
\label{xidefffxy}
\end{eqnarray}
where $p = (p_{\rm min},0)$, $p_{\rm min} \equiv 2 \pi/L$, and
$\widetilde{G}_o(q)$ is the Fourier transform of $G_o(x)$.  We also consider
some quantities that are invariant under RG transformations in the critical
limit, which we call renormalized couplings.  We consider the ratio $\xi/L$
and the quartic cumulants
\begin{equation}
U_{4}  \equiv { [ \rho_4 ]\over [\rho_2]^{2}}, \quad
U_{22} \equiv  {[ \rho_2^2 ]-[\rho_2]^2 \over [\rho_2]^2},
\label{uc}
\end{equation}
where $\rho_{k} \equiv \langle \; ( \sum_x q_x\; )^k \rangle$.

In the case of a $T=0$ transition with a nondegenerate ground state,
as expected in the 2D Ising glass model with a Gaussian disorder distribution,
we have $\chi\sim \xi^2$ for $T=0$, hence the corresponding
overlap-susceptibility exponent $\eta$ vanishes, $\eta=0$, and
\begin{equation}
U_{4}\to 1,\qquad U_{22}\to 0
\label{asu}
\end{equation}
for $T\to 0$. In particular, $U_{22}\to 0$ indicates the self-averaging of the
ground-state distribution, as already suggested by the results of
Ref.~\cite{LC-02}.  Moreover, since $\eta=0$, it is natural to conjecture that
the two-point overlap function becomes essentially Gaussian in the limit $T\to
0$. If this occurs, we also have $\xi/L\to\infty$.  As we shall see, the
results for the $\pm J$ Ising model are consistent with these predictions.

We also consider magnetic quantities. We define the magnetic susceptibility
$\chi_m$ and the second-moment correlation length $\xi_m$ in terms of the
magnetic two-point function
\begin{equation}
G_m(x) \equiv [\langle \sigma_0 \,\sigma_x \rangle].
\end{equation} 
For symmetric disorder distributions we have \cite{BY-86}
$\chi_m=1$ and $\xi_m=0$ for any $T$. For other values of $p$,
we expect them to converge to a finite nonuniversal value.
We also consider the four-point magnetic
susceptibility $\chi_{4m}$
defined by 
\begin{eqnarray}
&& \chi_{4m}\equiv [\mu_4 - 3 \mu_2^{2}]/L^2, \\
&& \mu_{k} \equiv \< \; ( \sum_x \sigma_x\; )^k \>.
\end{eqnarray}
For symmetric disorder distributions we have \cite{BY-86}
\begin{equation}
\chi_{4m}=4-6\chi.
\label{chi4mchio}
\end{equation}
Assuming universality we expect $\chi_{4m}\sim \chi$ also for nonsymmetric
disorder distributions, i.e., for any $p\ne 1/2$.

\section{Results at the glassy transitions}
\label{sec:results}

\begin{figure}
\includegraphics[width=20em,keepaspectratio]{rxivst0p8}
\vskip3mm
\includegraphics[width=20em,keepaspectratio]{u4vst0p8}
\vskip3mm
\includegraphics[width=20em,keepaspectratio]{u22vst0p8}
\caption{Phenomenological couplings $\xi/L$, $U_4$, and $U_{22}$ 
versus $T$ at $p=4/5$.}
\label{RvsT0.8}
\end{figure}

\begin{figure}
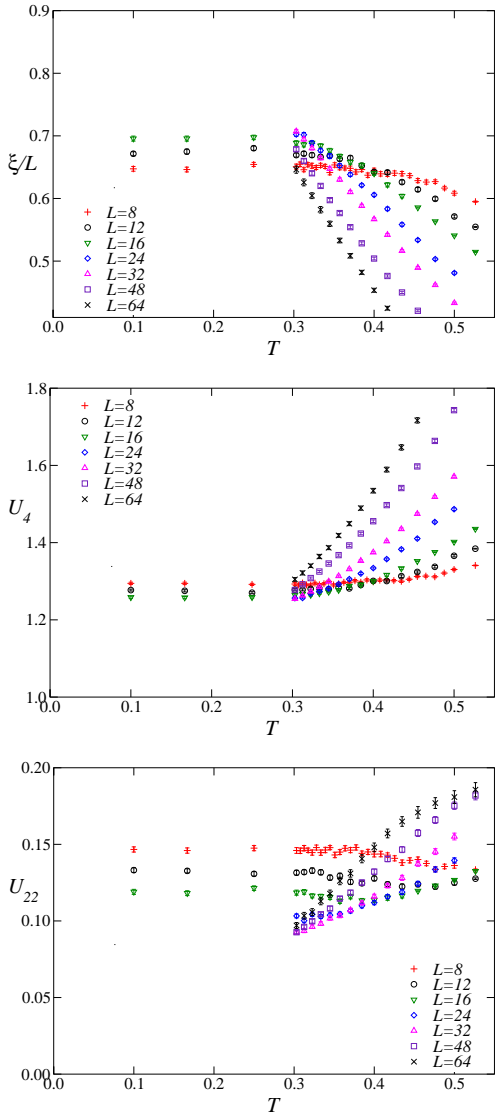

\includegraphics[width=20em,keepaspectratio]{rxivst0p5}
\vskip3mm
\includegraphics[width=20em,keepaspectratio]{u4vst0p5}
\vskip3mm
\includegraphics[width=20em,keepaspectratio]{u22vst0p5}
\caption{Phenomenological couplings $\xi/L$, $U_4$, and $U_{22}$ 
versus $T$ at $p=1/2$.}
\label{RvsT0.5}
\end{figure}

We perform MC simulations of the square-lattice $\pm J$ Ising model with
periodic boundary conditions for $p=4/5$ and $p=1/2$ and for several values of
the lattice size $L$, with $8\le L \le 64$.  We employ the Metropolis
algorithm, the random-exchange method~\cite{raex}, and multispin coding.
Furthermore, for the largest lattices we use the cluster algorithm described
in Ref.~\cite{Houdayer-01}.  For each lattice size we collect data in the
range $T_{\rm min}\le T \lesssim 1.1$.  At $p=4/5$ we take $T_{\rm min} =0.1$
for $L\le 32$, $T_{\rm min}=1/2.6\approx 0.38$ for $L=48$, $T_{\rm
  min}=1/2.7\approx 0.37$ for $L=64$.  At $p=1/2$ we take $T_{\rm min}=0.1$
for $L\le 16$ and $T_{\rm min}=1/3.3$ for $24\le L\le 64$.  Typically, we
consider $10^4$ disorder samples for each $T$ and $p$. In a few cases, we
consider $10^5$ disorder samples. In the following, we first discuss the
freezing regime, which occurs for sufficiently low temperatures in any finite
system, then we prove the universality of the glassy transition by considering
the FSS behavior of the renormalized couplings, and finally discuss the
behavior of the overlap susceptibility and of the magnetic quantities.

\subsection{The frozen regime}

In Figs.~\ref{RvsT0.8} and \ref{RvsT0.5} we show the MC estimates of 
$\xi/L$, $U_4$, and $U_{22}$.  
We note that the data corresponding to different lattice sizes 
cross each other around $T\approx 0.3$ and are mostly independent of 
$T$ for $T\lesssim 0.3$.  Similar results for
$U_4$ using the bimodal distribution were also reported in
Ref.~\cite{Houdayer-01}.  Usually, a crossing point 
corresponds to a transition point. Instead, in the present case in which the 
disorder variables are discrete, the crossing 
is due to a nonuniversal phenomenon which 
is related to the discreteness of the Hamiltonian spectrum
~\cite{KLC-07,JLMM-06}.

\begin{figure}
\includegraphics[width=20em,keepaspectratio]{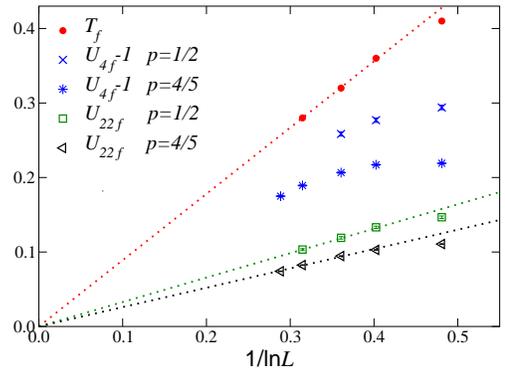}
\caption{Freezing temperature $T_f(L)$ as estimated from the onset of the 
   $T$-independent 
  behavior for $T\to 0$, and estimates of  $U_{22}$ and $U_4$ in 
  the frozen region (we indicate them by $U_{4f}$ and $U_{22f}$). Results for
  $8\le L \le 32$ and at the temperature $T=0.1$, which is
  well within the frozen region for the lattice sizes considered. The 
  dotted lines are drawn to guide the eye.}
\label{freezing}
\end{figure}

\begin{figure}
  \includegraphics[width=20em,keepaspectratio]{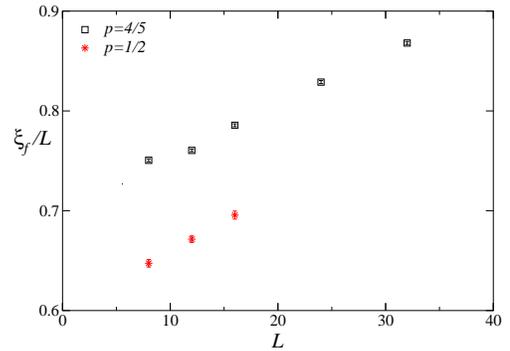}
\caption{ Estimates of $\xi_f/L$, where $\xi_f$ is the value of $\xi$ in the 
  frozen region.
}
\label{frozenrxi}
\end{figure}

To review the argument, 
let us consider the states corresponding to the two lowest energy
values for a given lattice size $L$. Their energies differ by $\Delta\equiv
E_1-E_0=4$ and their degeneracies are given by 
$N_0(L)$ and $N_1(L)$, respectively.  Numerical studies
\cite{LGMMR-04} have shown that $\ln N_1/N_0 \approx 4\ln L$.
At sufficiently low temperatures only the states with the lowest 
energy contribute to the thermodynamics. This occurs 
when $N_0(L)\gg N_1(L) e^{-\Delta/T}$, i.e., for 
\begin{equation} 
T \ll {\Delta\over \ln (N_1(L)/N_0(L))} \sim {1\over \ln L}.
\end{equation}
In this regime the observed behavior is independent of $T$.
In the opposite limit,
i.e., when $N_0(L)\ll N_1(L) e^{-\Delta/T}$, the presence of the
gap is negligible and the system is expected to have the same behavior as 
models with continuous distributions. The crossover from one regime to the 
other occurs at a $L$-dependent freezing temperature $T_f(L)$ which scales 
as $1/\ln L$. It is natural to define $T_f(L)$ by requiring 
$N_0(L)= N_1(L) e^{-\Delta/T_f}$, but this definition is somewhat 
unpractical. In practice, $T_f(L)$ can be estimated from the data
by identifying it with the temperature that marks the onset of the 
$T$-independent behavior of the different observables. For $p=4/5$, 
the estimates of $\xi/L$ and $U_4$ reported in Fig.~\ref{RvsT0.8} 
allow us to estimate
$T_f(L)\approx 0.41,\,0.36,\,0.32,\,0.28$ for $L=8,\,12,\,16,\,24$,
respectively. Slightly larger results are obtained by
using $U_{22}$.  The estimates of $T_f(L)$ for $p=1/2$ are close to 
those obtained for $p=4/5$, showing that $T_f(L)$ is
little dependent on $p$. Consistently with the above-reported argument, the
freezing temperature $T_f(L)$ approximately decreases as $1/\ln L$,
see Fig.~\ref{freezing}. A fit to $c/\ln L$ gives $c\approx 0.9$. 
In the frozen region the estimates of the renormalized couplings 
should be very close to the corresponding $T=0$ estimates, since they
are essentially determined by the lowest-energy configurations.
The data are consistent with this prediction. Indeed, 
see Fig.~\ref{freezing},
for both $p=4/5$ and $p=1/2$, $U_4$ and $U_{22}$ below $T_f$
slowly approach the values $U_{22}=0$ and
$U_4=1$, respectively, for $L\to\infty$.  In particular, $U_{22}$ apparently
vanishes as $U_{22}\sim 1/\ln L$. 
Below $T_f$ the ratio  $\xi/L$ increases
as $L$, see Fig.~\ref{frozenrxi}, indicating that $\xi \sim L^2$ at 
$T=0$. 
These results imply that, at $T=0$, the large-$L$ limit of the renormalized 
couplings is identical to that observed in models with continuous
distributions; in this case, as we discussed before, we predict 
$U_{22}\to 0$, $U_{4}\to 1$, and $\xi/L\to \infty$.
This equality should not be taken as an obvious fact.
For instance, the stiffness exponent is different in the two cases.

\subsection{Finite-size scaling in the presence of freezing}

The presence of freezing for $T<T_f(L)$
makes the study of the $T=0$ glassy critical behavior
quite hard.  Indeed, in order to observe the glassy critical behavior in Ising
glass models with a discrete distribution, one must approach $T=0$ by keeping
$T\gg T_f(L)$ for each lattice size. This makes a standard FSS analysis 
impossible. Indeed, in the FSS limit a RG invariant 
quantity $R$ should scale as 
\begin{equation}
R = f_R (TL^{1/\nu}).
\label{FSS1}
\end{equation}
The condition $T\gg T_f(L)$ implies that this scaling behavior can
only observed for 
\begin{equation}
T L^{1/\nu} \gg T_f(L) L^{1/\nu} \sim {L^{1/\nu}\over \ln L}.
\end{equation}
For $L\to\infty$, the ratio ${L^{1/\nu}/\ln L}$ diverges and thus this
makes the range of values of $T L^{1/\nu}$ which are accessible 
smaller and smaller as $L$ increases. This implies that the standard 
FSS limit, $T\to 0$, $L\to\infty$ at fixed $TL^{1/\nu}$ does not exist.
However, as we shall now discuss, one can still study FSS if one uses 
the ratio $\xi/L$ as basic FSS variable, i.e., if one considers the scaling 
form
\begin{equation}
R = g_R (\xi/L).
\label{FSS2}
\end{equation}
Usually, expressions (\ref{FSS1}) and (\ref{FSS2}) are equivalent. 
This is not the case here: only the FSS scaling form (\ref{FSS2})
holds in the presence of freezing. 
As is clear from Figs.~\ref{RvsT0.8} and \ref{RvsT0.5}, the ratio 
$\xi/L$ at fixed $L$ increases as $T$ decreases. Hence, the condition 
$T\gg T_f(L)$ translates into 
\begin{equation}
   {\xi\over L} \ll {\xi_f\over L},
\end{equation}
where $\xi_f$ is the value of $\xi$ in the frozen region. For $L\to \infty$
${\xi_f/L}$ diverges and thus, by increasing $L$, one has access to the 
whole FSS region. Thus, in the presence of freezing FSS can still be used 
but only in the form (\ref{FSS2}). Since $\nu$ does not appear in 
Eq.~(\ref{FSS2}), FSS cannot 
be used to determine $\nu$ (this exponent can only be determined by using 
infinite-volume data), though it can still be used to check universality.
Note that the presence of freezing and the limitations in the use of FSS
are always expected in models with a $T=0$ transition and discrete Hamiltonian
spectrum. In particular, these phenomena should also be considered in 
the 3D diluted $\pm J$ Ising model close to the percolation
point, where the glassy transition temperature vanishes
\cite{JR-08}. 

\subsection{Renormalized couplings and universality}

\begin{figure}
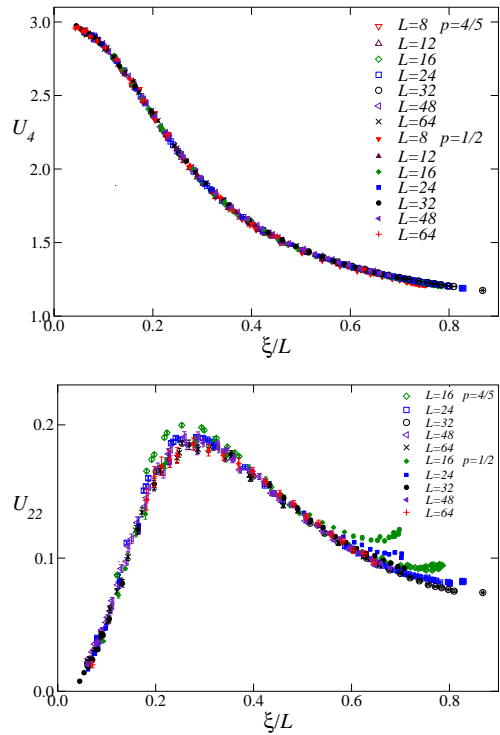

  \includegraphics[width=20em,keepaspectratio]{rxiu4} \vskip3mm
  \includegraphics[width=20em,keepaspectratio]{rxiu22}
\caption{Renormalized couplings $U_4$ and $U_{22}$ versus $\xi/L$ 
for $p=4/5$ and $p=1/2$. We only plot data satisfying $L\ge 16$ 
for clarity.
}
\label{scaling}
\end{figure}

\begin{figure}
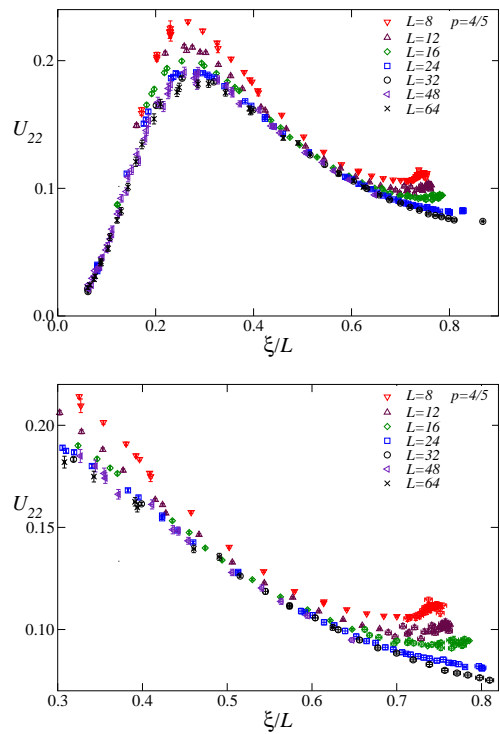

  \includegraphics[width=20em,keepaspectratio]{rxiu220p8} \vskip3mm
  \includegraphics[width=20em,keepaspectratio]{rxiu220p8tail}
\caption{Renormalized $U_{22}$ versus $\xi/L$ 
for $p=4/5$. We plot data for all available values of $L$. 
Below, we only show the
results for $\xi/L > 0.3$.
}
\label{scaling0p8}
\end{figure}

In order to verify universality, we consider the quartic cumulants $U_4$ and
$U_{22}$, which should scale according to Eq.~(\ref{FSS2}). The function
$g_R(x)$ should be universal, hence $p$-independent.  Universality is nicely
supported by the data shown in Fig.~\ref{scaling}.  As expected, we find that
$U_4\to 3$ and $U_{22}\to 0$ for $\xi/L\to 0$, and $U_4\to 1$ and $U_{22}\to
0$ for $\xi/L\to \infty$.  All data for $U_4$ fall onto a single curve with
small scaling corrections, supporting the existence of the finite-size scaling
limit in terms of $\xi/L$, as discussed in the previous section.  The
convergence to a single curve is also clear in the case of $U_{22}$, although
corrections are quite evident.  Note that, in the region around the peak, for
$\xi/L\approx 0.3$, the data at $p=4/5$ and $p=1/2$ converge from opposite
sides.  In these universality checks there are no free parameters to be
adjusted, and thus these comparisons provide strong support to the hypothesis
that these models belong to the same universality class.  An analogous
universality check was performed in Ref.~\cite{KLC-07}. By studying the FSS
behavior of $U_4$ versus $\xi/L$, it provided
numerical evidence that the $\pm J$ Ising model
with a bimodal distribution (our model with $p=1/2$) and that with a Gaussian
distribution of the couplings belong to the same universality class.  It is
interesting to note that while the data for $U_4$ scale nicely up to $\xi/L
\lesssim 0.9$, for $U_{22}$ significant deviations are observed at 
smaller values of $\xi/L$, see Fig.~\ref{scaling0p8}.  For $p=4/5$ and $L=8$
the data show significant deviations close to the peak, then approach the
common curve and then show again a significant deviation --- the data turn up
--- for $\xi/L > (\xi/L)_{\rm max} \approx 0.55$. A similar phenomenon occurs
for $L=12$. For $L=16$ deviations close to the peak are quite small, but again
the data begin to turn up as $\xi/L > (\xi/L)_{\rm max} \approx 0.6$.  For
$L=24$ FSS holds quite nicely, at least up to $(\xi/L)_{\rm max} \approx
0.65$. The value $(\xi/L)_{\rm max}$ marks the onset of the crossover region
between the critical regime where FSS holds and the freezing regime that sets
in at $\xi_f/L$. Note that $\xi_f/L$ is significantly larger than
$(\xi/L)_{\rm max}$, indicating that the breaking of FSS occurs before
freezing. For $p=1/2$ the conclusions are similar, although, for a given $L$,
$(\xi/L)_{\rm max}$ is significantly smaller than the corresponding value for
$p=4/5$. For instance, for $L=16$, we have $(\xi/L)_{\rm max} \approx 0.45$
for $p=1/2$ and $(\xi/L)_{\rm max} \approx 0.60$ for $p=4/5$. This clearly
reflects the fact that $\xi_f$ for $p=1/2$ is smaller than for $p=4/5$, see
Fig.~\ref{frozenrxi}.

\subsection{Overlap susceptibility and exponent $\eta$}

Finally, we investigated the critical behavior of the overlap susceptibility.
As discussed in Ref.~\cite{HPV-08} it should behave in the FSS limit as 
\begin{eqnarray}
  \chi = \overline{u}_h^2(T) L^{2-\eta} F_\chi(\xi/L),
\label{ansatz-chi}
\end{eqnarray}
where $\overline{u}_h$ is an analytic function of $T$ which is 
related to the overlap-magnetic scaling field.
In order to determine $\eta$ we have performed fits of the data 
with $p=4/5$ to 
\begin{eqnarray}
\ln \chi = (2-\eta)\ln L + P_n(T) + Q_m(\xi/L),
\label{fitchi}
\end{eqnarray}
where $P_n(x)$  and $Q_m(x)$ are polynomials in $x$ with $P_n(0) = 0$.
To avoid any bias from the presence of the freezing region, we have only
used the data satisfying $\xi/L < (\xi/L)_{\rm max}$, where, for $L\le 24$, 
$(\xi/L)_{\rm max}$ is the value determined before from the analysis of 
$U_{22}$. For $L=32$ we used somewhat arbitrarily
$(\xi/L)_{\rm max} = 0.7$, while for $L=48,64$ we used all our data 
which in any case satisfy  $\xi/L < 0.65$. To identify scaling corrections
we only considered data satisfying $T<T_{\rm max}$ and $L> L_{\rm min}$
for several values of $T_{\rm max}$ and $L_{\rm min}$. 
The results depend strongly on these parameters. For $T_{\rm max} = 1.2$
we obtain $\eta = 0.39(1)$, 0.33(3) for $L_{\rm min} = 8,16$. 
For $L_{\rm min} = 16$, we obtain $\eta = 0.27(3), 0.24(3), 0.22(4)$ 
for $T_{\rm max} = 1,0.8,0.6$. Apparently the results always 
decrease as $T_{\rm max}$ decreases and $L_{\rm min}$ increases. 
It is impossible to estimate reliably $\eta$ from these results and 
thus we only quote an upper bound: 
\begin{equation}
\eta \lesssim 0.2~.
\end{equation}
In principle, the same analysis can be performed for $p=1/2$. However, in
this case the estimates of $(\xi/L)_{\rm max}$ are smaller, so that fewer 
data can be used in the fit. In practice, no estimates of $\eta$ can be 
obtained.

\begin{figure}
\includegraphics[width=20em,keepaspectratio,angle=0]{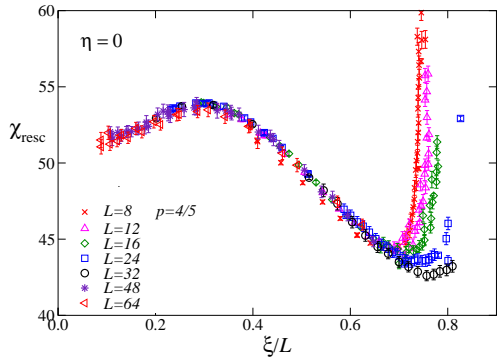}
\caption{Rescaled susceptibility $\chi_{\rm resc}$ defined in 
Eq.~(\ref{chiresc}) for $p=4/5$ and $\eta = 0$.}
\label{plot-chiresc}
\end{figure}

\begin{figure}
\includegraphics[width=20em,keepaspectratio,angle=0]{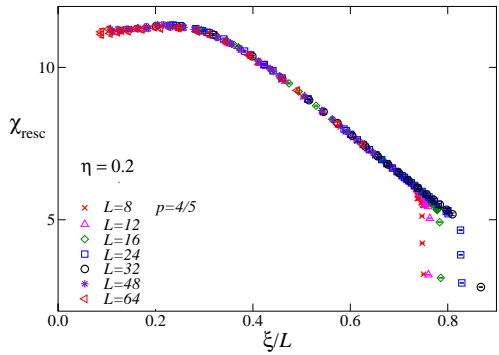}
\caption{Rescaled susceptibility $\chi_{\rm resc}$ defined 
in Eq.~(\ref{chiresc}) for $p=4/5$ and $\eta = 0.2$.}
\label{plot-chiresc2}
\end{figure}

As we mentioned at the beginning there are strong theoretical reasons to 
expect $\eta = 0$. Thus, we tried to verify whether 
our results are consistent with this hypothesis. 
For this purpose we considered the data with $T_{\rm max} = 1.0$
and $L_{\rm min} = 12$ and we fitted them to Eq.~(\ref{fitchi})
setting $\eta = 0$. We find that the data satisfying 
$\xi/L < (\xi/L)_{\rm max}$ are reasonably described by Ansatz
(\ref{ansatz-chi}). In Fig.~\ref{plot-chiresc} we report
\begin{equation}
\chi_{\rm resc} = \chi L^{-2} e^{-P_n(T)},
\label{chiresc}
\end{equation}
where $P_n(T)$ is the polynomial determined in the fit (\ref{fitchi}). 
The agreement is quite good up to $(\xi/L)\approx 0.65$. It should be noted that, 
if we also include data with $\xi/L > (\xi/L)_{\rm max}$, the fits 
become much less dependent on $T_{\rm max}$ and $L_{\rm min}$, 
exclude $\eta= 0$, and give the estimate 
$\eta \approx 0.2$: Apparently the data that belong to the 
region $ (\xi/L)_{\rm max} < \xi/L < \xi_f/L$ are well described 
by Eq.~(\ref{ansatz-chi}) with $\eta = 0.2$, while they cannot be fitted 
by taking $\eta = 0$. 
This is evident from Fig.~\ref{plot-chiresc2} where 
we report $\chi L^{-1.8} \exp[-P_n(T)]$: in this figure essentially 
all data fall on a single rescaled curve. Note that the quality of the 
collapse is only marginally better than that given in Fig.~\ref{plot-chiresc}
--- the $\chi^2/{\rm DOF}$ of the fit (DOF is the number of degrees of 
freedom of the fit) is similar in the two cases ---
although this is not evident from the figures  
since they have a completely different vertical scale. 

\begin{figure}
\includegraphics[width=20em,keepaspectratio,angle=0]{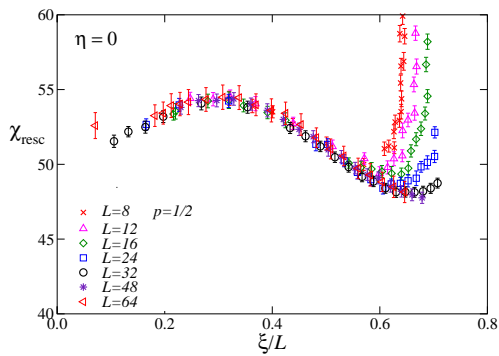}
\caption{Rescaled susceptibility $\chi_{\rm resc}$ defined in 
Eq.~(\ref{chiresc}) for $p=1/2$ and $\eta = 0$.}
\label{plot-chiresc3}
\end{figure}

As a final check we verify if the data at $p=0.5$
are also consistent with $\eta = 0$. The results are reported in
Fig.~\ref{plot-chiresc3}. Also in this case 
the data with $\xi/L \lesssim 0.65$ are consistent with $\eta = 0$. 
In the plot we have also multiplied $\chi_{\rm resc}$ by a constant
in such a way that 
$\chi_{\rm resc}$ assumes the value $\chi_{\rm resc} \approx 56$ for 
$\xi/L \approx 0.3$ as it does for $p = 4/5$. With this choice the 
curves for $\chi_{\rm resc}$ should be the same for both values of $p$. 
As it can be seen the shape of the two curves is indeed the same. 
Quantitatively, the two curves are the same up to $\xi/L \approx 0.45$, 
while they differ significantly for 
$\xi/L \approx 0.6, 0.7$. This is not surprising. As we have 
explained, for $p=1/2$, the data such that $\xi/L\gtrsim 0.5$ are probably
already in the crossover region before the onset of freezing. 

\subsection{Magnetic quantities}

\begin{figure}
\vskip5mm
\includegraphics[width=20em,keepaspectratio]{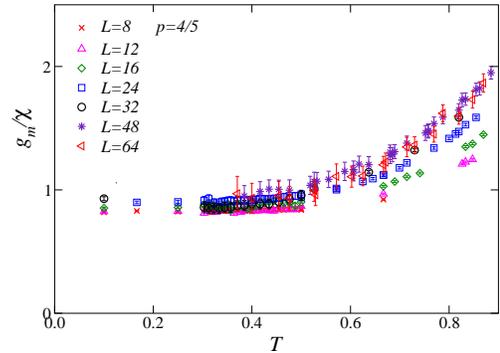}
\caption{The ratio $g_{m}/\chi$ at $p=4/5$.}
\label{g4}
\end{figure}

The magnetic variables do not become critical in limit $T\to 0$. Indeed, the
magnetic susceptibility $\chi_m$ and second moment correlation length $\xi_m$
are finite in the limit $T\to 0$. For 
$p=1/2$ we have $\chi_m=1$ and $\xi_m=0$ for any $T$. For other values of $p$,
they converge to nonuniversal values such that 
$\chi_m>1$ and $\xi_m>0$. For $p=4/5$ we find $\chi_m\approx 24$ and
$\xi_m\approx 3.0$ in the limit $T\to 0$.  We also consider 
the four-point magnetic
susceptibility $\chi_{4m}$ which should scale as 
$\chi$ in the critical limit. In Fig.~\ref{g4} we plot the ratio 
$g_{m}/\chi$, where 
$g_{m}\equiv - \chi_{4m}/(\chi_m^2\xi_m^2)$. This quantity shows 
smaller scaling corrections than $\chi_{4m}/\chi$ and clearly converges to
an $L$ independent constant. The asymptotic behavior sets in for 
$T\lesssim 0.5$, before the 
freezing region $T\lesssim 0.35$.

\section{The overlap critical behavior along the low-temperature
paramagnetic-ferromagnetic transition line}
\label{sec:resultsPF}

We now investigate the behavior of the overlap correlations along the
low-temperature PF transition line, 
see Fig.~\ref{phasedia}, from the MNP to the $T=0$
axis.  The critical behavior of the magnetic correlations was 
numerically studied in Ref.~\cite{PTPV-09} 
by varying $p$ for two values of $T$: $T=1/1.55$ and $T=1/2$. 
It was found that the critical behavior was universal, controlled by a 
single strong-disorder fixed point, 
with critical exponents $\nu\approx 3/2$ and $\eta_m\approx 1/8$.  

The ferromagnetic
$T=0$ transition point at $p=p_0\approx 0.897$, where the low-temperature PF
transition line ends, is a multicritical glassy point (MGP), 
because it is connected to
three phases and it is the intersection of two different transition lines, the
PF line at $T>0$ and the glassy line at $T=0$.  At $T=0$ the critical point at
$p=p_0$ separates a ferromagnetic phase from a $T=0$ glassy phase, while for
$T > 0$ the transition line separates a ferromagnetic from a paramagnetic
phase.  Therefore, on general grounds, the critical behavior at $T=0$ and 
$p=p_0$ should differ both from that observed along the PF line and that 
observed along the glassy line $T=0$, $p>p_0$, unless
the magnetic and glassy critical modes are effectively decoupled.
Such a decoupling is apparently supported by the 
numerical results of Refs.~\cite{McMillan-84,WHP-03,AH-04,PHP-06}.
Indeed, the estimates of magnetic critical exponents at $T=0$
are quite close and
substantially consistent with those found along the PF transition line below
the MNP.  All results are therefore consistent with a single magnetic fixed
point that controls the magnetic critical behavior both at $T > 0$ and at
$T=0$.  The analysis of the overlap correlation functions also supports
the decoupling of the critical modes. Indeed, the arguments we gave 
in Sec.~\ref{sec:mc} on the behavior of the (overlap) renormalized couplings
and correlation functions 
should hold at $T=0$ for any value of $p$ --- hence, in the ferromagnetic phase,
at the MGP as well as along the glassy transition line --- since they only rely 
on the assumption of a nondegenerate ground state. 
Therefore, also at the $T=0$ MGP we expect
$U_4 = 1$, $U_{22} = 0$, $\xi/L=\infty$, and $\eta = 0$. 
The behavior of the overlap correlations is therefore identical 
for $p>p_0$, for $p < p_0$, and at the MGP, in agreement with the decoupling 
scenario.

We wish now to understand the critical behavior of the overlap correlations 
along the PF line. 
In order to investigate this issue, we perform MC simulations at two
critical points along the low-temperature PF line, at $[T=1/1.55\approx
0.645,\,p=0.8915(2)]$ and $[T=1/2,\,p=0.8925(1)]$, as determined in
Ref.~\cite{PTPV-09}, for lattice sizes up to $L=48$.  The 
FSS analysis of the data of the overlap susceptibility $\chi$ shows that
$\chi\sim L^{2-\eta}$ with quite small but nonzero values of $\eta$.
Fits of the data satisfying $L\ge 16$ which take into account 
the nonanalytic scaling corrections give
$\eta=0.046(6)$ at $T=1/1.55$ and $\eta=0.038(4)$ at 
$T=1/2$. Note
that these values are much smaller than the pure Ising value $\eta=1/2$
(which is simply twice the value of the magnetic exponent $\eta_m = 1/4$)
holding along the PF line from the pure Ising point to the MNP, 
and also much
smaller than the value $\eta=\eta_m=0.177(2)$ at the MNP point.  The FSS analyses
of the renormalized couplings $\xi/L$, $U_4$ and $U_{22}$ lead to
apparently $T$-dependent critical values: $U_{22}=0.018(2)$, $U_4=1.044(4)$,
$\xi/L=1.7(1)$ at $T=1/1.55$, and $U_{22}=0.008(2)$, $U_4=1.025(2)$,
$\xi/L=2.2(1)$ at $T=1/2$ (where the errors are essentially due to the
uncertainty on $p_c$ and on the scaling correction exponent).  
Note that these values are very close to the values at the $T=0$ MGP;
still we consider unlikely that the overlap behavior is the 
same as that occurring at the MGP, mainly because this would require 
$\xi/L = \infty$ along the whole PF line between the MNP and the MGP. 
Indeed, 
the condition $\xi/L = \infty$ is quite unlikely for a finite-temperature 
transition.  
These results can be better explained 
by a $T$-dependent asymptotic critical behavior 
which, with decreasing $T$, approaches the $T=0$
glassy behavior characterized by the values $\eta=0$, $U_{22}=0$,
$U_4=1$, $\xi/L = \infty$. 

\section{Conclusions} \label{sec:conclusions}

In this paper we consider the two-dimensional $\pm J$ Ising model, 
focusing mainly on the $T=0$ glassy transition occurring for 
$1-p_0 < p < p_0$. 
The main results are the following.
\begin{itemize}
\item[i)] 
We first discuss the freezing phenomenon that occurs on any finite lattice
at sufficiently low temperatures. We investigate 
the behavior of several quantities in this regime, verifying
explicitly the expected logarithmic dependence on the lattice size $L$. 
We also show that 
the presence of this regime makes it impossible to use the standard form of 
FSS: the FSS limit $T\to 0$, $L\to \infty$ at fixed $TL^{1/\nu}$ does not 
exist. FSS can be formulated only if one considers $\xi/L$ as basic FSS 
variable. 
\item[ii)] We study  the FSS behavior of the renormalized couplings
$U_{4}$ and $U_{22}$ for $p=4/5$ and $p=1/2$ as a function of $\xi/L$. 
We find that they have the same FSS curves, 
a clear indication that the
critical behavior for $p=4/5$ and $p=1/2$ is the same. This allows us 
to conjecture that the critical behavior is independent of $p$ in the interval
$1-p_0 < p < p_0$, and therefore,
together with the results of Ref.~\cite{KLC-07}, that there exists a 
single universality class for 2D Ising glassy transitions. We also 
investigate in detail the critical behavior of the overlap susceptibility,
showing that the numerical data are consistent with $\eta = 0$, 
if one discards data that are close to the region where freezing occurs.
\item[iii)]
Finally, we discuss the critical behavior of the overlap variables along the 
PF line. An analysis of the numerical data available for the $T=0$ MGP
indicates that at this point glassy and magnetic modes are decoupled. 
For $T>0$ we observe an apparent $T$-dependent critical behavior. 
Note that a similar 
phenomenon was also observed in the XY model with random shifts
\cite{APV-10}: along 
the critical line that starts at the XY pure point and ends at the 
Nishimori multicritical point, magnetic quantities show a universal behavior
while overlap variables show a disorder-dependent critical behavior.
\end{itemize}


\begin{thebibliography}{99}
\bibitem{Nline} The Nishimori line is defined by the equation $T=T_N(p) =
  2/[\ln p - \ln(1-p)]$. Along this line several rigorous results can be
  proved~\cite{Nishimori-81,Nishimori-01}, such as the equality of the
  magnetic and overlap two-point correlations.
\end{thebibliography}
\end{document}